\documentclass[]{aa}

\usepackage{amsmath}
\usepackage{graphicx}

\usepackage{natbib}
\bibpunct{(}{)}{;}{a}{}{,}

\usepackage{txfonts}

\begin{document}

\title{Lagrangian coherent structures in photospheric flows and\\
 their implications for coronal magnetic structure}

\titlerunning{Photospheric flows and coronal magnetic structure}

\author{A.~R. Yeates\inst{\ref{inst0}}
\thanks{Previously at Division of Mathematics, University of Dundee}
	\and
	G. Hornig\inst{\ref{inst1}}
	\and
	B.~T. Welsch\inst{\ref{inst2}}
	}
\institute{Department of Mathematical Sciences, Durham University, Durham, DH1 3LE, UK,
\email{anthony.yeates@durham.ac.uk}\label{inst0}
\and
Division of Mathematics,
University of Dundee, Dundee, DD1 4HN, UK,
\email{gunnar@maths.dundee.ac.uk}\label{inst1}
\and
Space Sciences Laboratory, University of California, Berkeley, CA 94720, USA,
\email{welsch@ssl.berkeley.edu}\label{inst2}
}

\date{Received (date) / Accepted (date)}

\abstract
{}
{
We show how the build-up of magnetic gradients in the Sun's corona may be inferred directly from photospheric velocity data. This enables  computation of magnetic connectivity measures such as the squashing factor without recourse to magnetic field extrapolation.
}
{Assuming an ideal evolution in the corona, and an initially uniform magnetic field, the subsequent field line mapping is computed by integrating trajectories of the (time-dependent) horizontal photospheric velocity field. The method is applied to a 12 hour high-resolution sequence of photospheric flows derived from Hinode/SOT magnetograms.}
{We find the generation of a network of quasi-separatrix layers in the magnetic field, which correspond to Lagrangian coherent structures in the photospheric velocity. The visual pattern of these structures arises primarily from the diverging part of the photospheric flow, hiding the effect of the rotational flow component: this is demonstrated by a simple analytical model of photospheric convection. We separate the diverging and rotational components from the observed flow and show qualitative agreement with purely diverging and rotational models respectively. Increasing the flow speeds in the model suggests that our observational results are likely to give a lower bound for the rate at which magnetic gradients are built up by real photospheric flows. Finally, we  construct a hypothetical magnetic field with the inferred topology, that can be used for future investigations of reconnection and energy release.}
{}

\keywords{Magnetic fields - Sun: photosphere - Sun: corona - Sun: magnetic topology}

\maketitle

\section{Introduction}

This paper proposes a straightforward method to study the time-dependent build up of structure in the Sun's coronal magnetic field, based on observations of horizontal velocity fields in the solar photosphere. The ultimate objective is to determine whether the energy built up and released in the coronal magnetic field as a result of photospheric convection is sufficient to heat the corona via the Parker mechanism \citep{parker1972}. In this theory, footpoint braiding generates localised magnetic gradients and thin current sheets in the corona, leading to ubiquitous reconnection. There are two requirements that must be satisfied by the observed motions if this is to work: (1) they must have a tendency to create the required magnetic gradients, and (2) they must do so quickly enough to generate a sufficient overall reconnection rate. The method proposed here aims to investigate these questions in a practical way, given currently available observations.

Numerous studies have found that the topology and connectivity of the coronal magnetic field play a primary role in determining when and where magnetic reconnection will take place \citep{birn2007}. An increasingly popular means to characterise the connectivity of 3D magnetic fields has been through the so-called \emph{squashing factor} $Q$ of the magnetic field line mapping \citep{titov2002}. Quasi-separatrix layers (QSLs),where $Q$ is high, represent locations in the magnetic field with large gradients in field line connectivity. They have been identified with the locations of flares and X-ray bright points \citep{demoulin1996,demoulin1997,mandrini1996,gaizauskas1998,wang2000}. Magnetic reconnection is suggested to occur preferentially at QSLs \citep{demoulin2006,santos2008}.

Although $Q$ is defined solely by the connectivity of field lines (the mapping between photospheric footpoints), existing studies have all calculated $Q$ by first constructing a 3D magnetic field, then tracing field lines to determine the mapping. Our method determines the field line mapping, and hence $Q$, in a fundamentally different way. Given some initial field line mapping at time $t_0$, and assuming an ideal evolution of the coronal magnetic field, the field line mapping at a later time depends only on the sequence of photospheric footpoint motions. We propose to take advantage of this fact to compute the field line mapping directly from photospheric velocity data.

The main advantage of our proposed method is that it avoids the need to extrapolate a 3D magnetic field from photospheric magnetograms. This is problematic for studies of magnetic topology because standard techniques (such as potential fields) applied to a sequence of photospheric magnetograms will not give the correct field line topology commensurate with an ideal evolution from one time to the next. To avoid this problem, a number of studies have used time-dependent 3D simulations to model how topological structure develops in response to simple boundary motions \citep{milano1999,galsgaard2003,aulanier2005,masson2009} or to boundary motions derived directly from observed magnetograms \citep{mackay2011}. But such time-dependent simulations inevitably suffer from numerical dissipation, leading to inaccuracies in magnetic topology. Our method avoids this.

The main limitation of our proposed method is that it cannot determine the initial field line mapping at time $t_0$, only that resulting from subsequent footpoint motions. The initial mapping will likely contain pre-existing magnetic structure and QSLs. Unfortunately, existing extrapolation techniques cannot uniquely determine this initial magnetic topology from available magnetogram observations. Note that this limitation is shared by all of the studies cited in the previous paragraph, which must also assume some initial magnetic field; typically a potential field extrapolation is used. A similar limitation applies to studies of magnetic helicity using photospheric flows and magnetograms \citep{demoulin2009}. In this paper, we simply assume a uniform initial field, whose field line mapping is the identity. This is the most conservative choice: $Q$ values derived from the subsequent mappings will almost certainly be a lower bound for the real coronal magnetic field.

\begin{figure}
\centering
\includegraphics[width=\columnwidth]{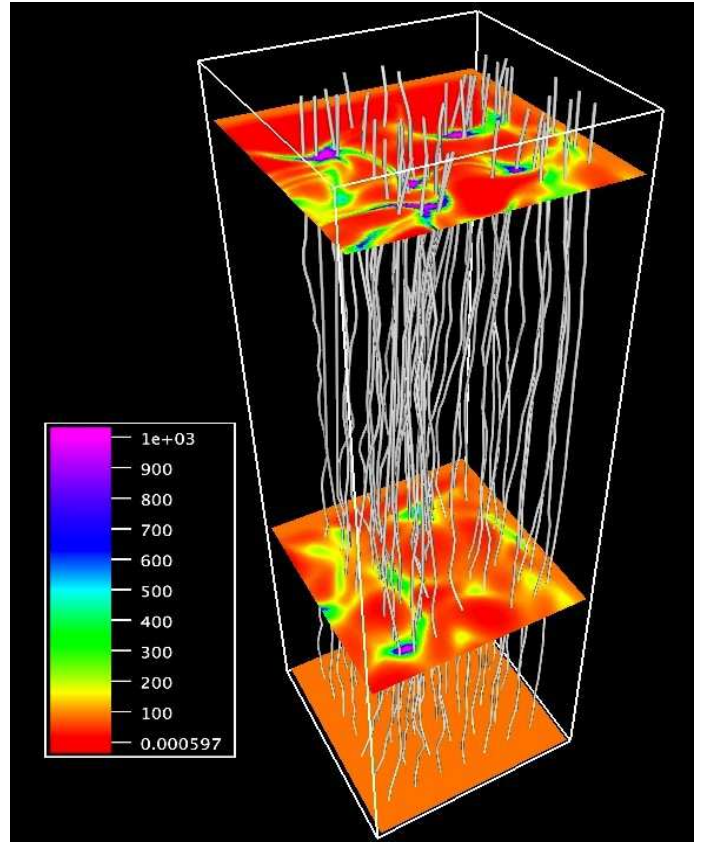}
\caption{A hypothetical magnetic field (Appendix \ref{sec:mag}) whose field line mapping matches that inferred from the observed velocity field. The magnetic field lines in this particular field are simply trajectories of tracer particles in the 2D, time-dependent velocity field (with time increasing vertically). Contour slices show $B_z$ at different $z$ with the colour scale in gauss: the initial distribution (on the lower boundary) is a uniform field $B_z(x,y,0)=88\,\textrm{G}$, which was the average field strength in this region of the original SOT magnetogram.}
\label{fig:suspend}
\end{figure}

An advantage of taking the initial field to be uniform is that one may construct a hypothetical magnetic field with the correct topology simply by taking the magnetic field lines to be the trajectories of the velocity field. Examples of such trajectories are shown in Fig. \ref{fig:suspend}, where time increases vertically. The cross-sections show the magnitude of $B_z$ for this hypothetical magnetic field; these may be thought of as the photospheric magnetograms at subsequent times. (See Appendix \ref{sec:mag} for details of the calculation.) The concentration of magnetic flux in a ``network'' of convective cell boundaries is evident. Of course, this hypothetical field does not represent a realistic extrapolation of the magnetic field that would be seen on the Sun, since it starts from a uniform field at $t_0$, and only takes into account photospheric motions at one end of the field lines. Rather, we envisage using it as the starting point for 3D MHD simulations investigating energy release, with the advantage of having determined accurately the change in field
line mapping.

To demonstrate the proposed method, we apply it to a 12 hour sequence of photospheric velocities derived by local correlation tracking in Hinode/SOT \citep[Solar Optical Telescope,][]{tsuneta2008} magnetograms. The observed velocities are described in Sect. \ref{sec:data}, while the inferred magnetic field line mapping is presented in Sect. \ref{sec:map}. In addition to $Q$, we compute the \emph{finite-time Lyapunov exponent} (FTLE) field $\sigma$. This measure is a popular method in fluid mechanics for identifying so-called Lagrangian Coherent Structures (LCSs) in velocity fields. Like $Q$, $\sigma$ measures the maximum separation rate of initially nearby trajectories, and we illustrate how QSLs in the field line mapping correspond to LCSs in the photospheric velocity field. In Sect. \ref{sec:interp} we explain the pattern observed in the $Q$ or $\sigma$ fields using a simple analytical model of photospheric convection. By varying the model parameters, we predict how the field line mapping would be expected to change given observations at higher resolution of faster flows. Conclusions are given in Sect. \ref{sec:conclusions}.

\section{Photospheric Velocity Data} \label{sec:data}

Our velocity data have been derived by local correlation tracking in magnetograms, although the method could be applied to velocity fields from any source. Detailed analysis of the data reduction procedure is given by \citet{welsch2011}. Briefly, we use Stokes $V/I$ from Hinode/NFI (Narrowband Filter Imager) observations in Fe I $6302\AA$ of active region 10930. These were calibrated to gauss following Equation (1) of \citet{isobe2007}, and the noise level estimated at $\sim 17\,\textrm{G}$ by fitting the core of histogrammed field strengths \citep{hagenaar1999}. In view of the subsequent reduction of noise by averaging in the tracking procedure, a tracking threshold of $15\,\textrm{G}$ was chosen, with no velocities assigned to pixels below this threshold. The magnetogram pixels are binned (2x2) from $0.16''$ to $0.32''$, consistent with SOT's $0.3''$ diffraction limit at this wavelength. The cadence of the images is $\sim 121\,\textrm{s}$, and the sequence runs from 14:00UT on 12 December to 02:58UT on 13 December 2006.

The velocity field is extracted from the magnetograms using the Fourier local correlation tracking (FLCT) method \citep{welsch2004,fisher2008}. The method has a number of parameters: optimum values have been determined by an autocorrelation analysis, aiming to maximise frame-to-frame correlations and ensure robustness in the velocity estimate \citep[see][]{welsch2011}. Here, the windowing/apodization parameter is set to 4 to avoid too much spatial averaging of small-scale flows. The sampling time between subsequent frames is chosen as $\Delta t = 8\,\textrm{mins}$. This is small enough to avoid significant decorrelation, but large enough to allow for boxcar averaging of 5 magnetograms to produce each frame, which greatly reduces noise. We have repeated the calculations with $\Delta t=4\,\textrm{mins}$ with qualitatively similar results.

\begin{figure}
\centering
\includegraphics[width=\columnwidth]{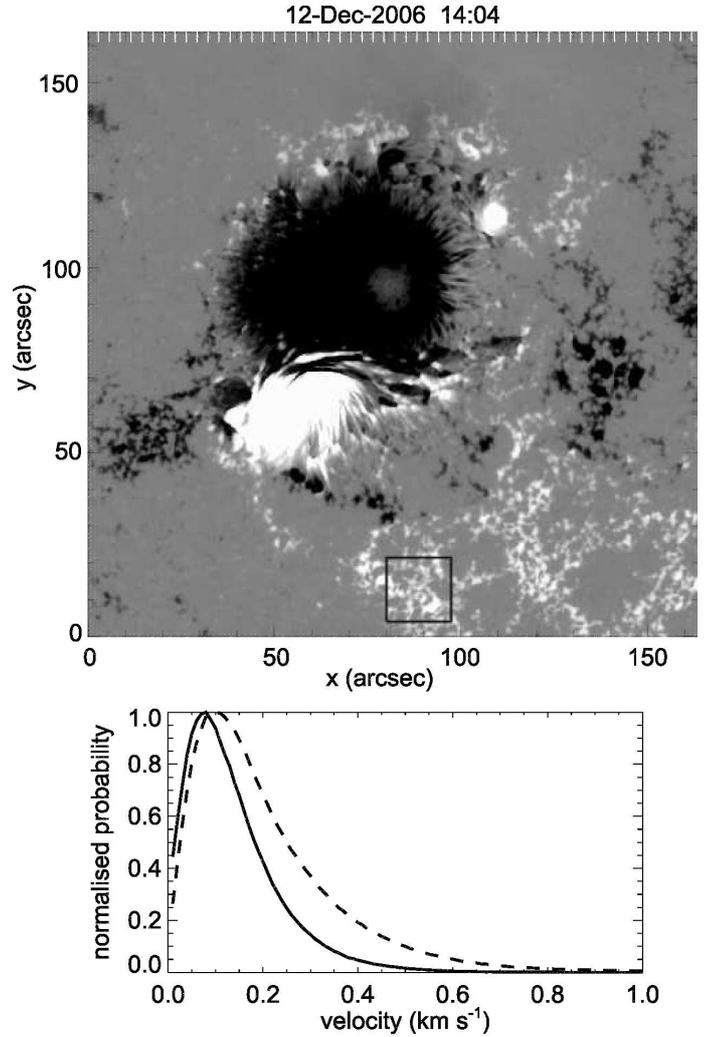}
\caption{{\it Left:} Location of the $12.4\,\textrm{Mm}\times 12.4\,\textrm{Mm}$ analysis region (black box) in the full SOT line-of-sight magnetogram ($1''=725\,\textrm{km}$). {\it Right:} Histogram of measured $|{\bf v}|$ in the analysis region, over the entire 12 hour dataset. The solid line shows the filtered data, while the dashed line shows the original unfiltered data. Each curve is normalised by its own maximum.}
\label{fig:location}
\end{figure}

For the analysis in this paper, we select a unipolar plage region of size $12.4\,\textrm{Mm}\times 12.4\,\textrm{Mm}$ (approximately $17''\times 17''$), away from the main sunspots, as shown in Fig. \ref{fig:location} (left). This is to avoid the large-scale flow associated with emerging flux and rotation of the sunspots. Since the magnetic flux in our region is concentrated in the supergranular lanes, there are inevitably areas where the line-of-sight magnetic field is too weak for reliable estimation of the velocity. This particular region has been chosen to minimise this problem over the length of the time sequence, although there are several regions where the velocity suffers locally from high-frequency noise. We have removed this noise with minimal disturbance to the well-resolved regions by applying a low-pass (Butterworth) filter to the velocity fields in Fourier space. Histograms of the velocities both with and without filtering are shown in Fig. \ref{fig:location} (right). The mean flow speed is of the order $0.1\,\textrm{km}\,\textrm{s}^{-1}$, which is rather lower than reported speeds for granular flows \citep[$\sim 1\,\textrm{km}\,\textrm{s}^{-1}$,][]{rieutord2010}. There are a number of possible reasons for this. Firstly, there is a likely averaging effect due to the convective cells being close to our spatial resolution of $0.3''$. In addition, comparative tests show that FLCT has a bias toward underestimating speeds \citep{welsch2007}. However, it should be noted that the FLCT method tracks coherent magnetic features, which are expected to move more slowly than surrounding plasma due to suppression of convection \citep{title1992,berger1998}. The possible effect of faster flows is explored in Sect. \ref{sec:interp}.

\begin{figure}
\includegraphics[width=\columnwidth]{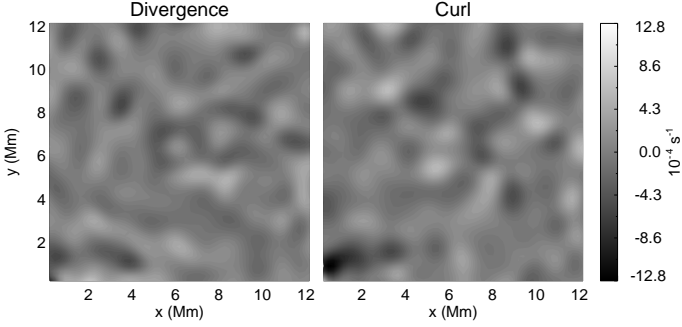}
\caption{Snapshots of $\Delta$ ({\it left}) and $\omega$ ({\it right}) in the analysis region, at 14:04UT. The grey scale extends to the maximum absolute value of each quantity.}
\label{fig:divcurl}
\end{figure}

Figure \ref{fig:divcurl} shows snapshots of the horizontal divergence of the (filtered) velocity,
\begin{equation}
\Delta\equiv\nabla\cdot{\bf v}=\partial v_x/\partial x + \partial v_y/\partial y,
\end{equation}
and vertical component of the vorticity,
\begin{equation}
\omega\equiv{\bf e}_z\cdot(\nabla\times{\bf v})= \partial v_y/\partial x - \partial v_x/\partial y.
\end{equation}
The root-mean-square (RMS) divergence over the whole 12 hour dataset is $2.06\times 10^{-4}\,\textrm{s}^{-1}$ (or $3.49\times 10^{-4}\,\textrm{s}^{-1}$ before filtering), while the RMS $\omega$ is $1.97\times 10^{-4}\,\textrm{s}^{-1}$ ($3.16\times 10^{-4}\,\textrm{s}^{-1}$ before filtering). Thus the divergence and curl of the velocity field are comparable in magnitude. This will be important for understanding the inferred magnetic field line mapping.

\section{Inferred Magnetic Field Line Mapping} \label{sec:map}

The field line mapping is simply given by following trajectories (particle paths) in the observed time-dependent 2D velocity field. So a particle starting from $(x_0,y_0)$ at time $t_0$ is mapped to ${\bf f}(x_0,y_0,t_1)=\big(f_x(x_0,y_0,t_1),f_y(x_0,y_0,t_1) \big)$ at time $t_1$, where ${\bf f}$ is found by integrating
\begin{equation}
\frac{d{\bf f}(x_0,y_0,t)}{d t} = {\bf v}\big({\bf f}(x_0,y_0,t)\big)
\label{eqn:traj}
\end{equation}
from $t=t_0$ to $t=t_1$, for which we use a second-order method. The observed velocity fields are interpolated using a local tricubic method \citep{lekien2005} that also gives continuous first derivatives of ${\bf v}$. Linear interpolation does not produce smooth enough trajectories for computing the Lagrangian structures we are interested in. Equation \eqref{eqn:traj} is integrated on a grid of $864\times 864$ starting points to give the 2D field line mapping at a given end-time $t_1$. The high resolution is needed to accurately determine the $Q$ and $\sigma$ fields, which typically vary on a smaller scale than the velocity field itself \citep[see][]{shadden2011}.

To analyse the resulting time sequence of mappings for different $t_1$, we compute two measures of the mapping gradient: the squashing factor $Q$ and the FTLE field $\sigma$. The former is frequently used to characterise 3D magnetic field structure, while the latter is used to characterise particle paths in time-dependent 2D velocity fields. In fact, both are rather similar measures of the local rate of stretching at a given point, and both are defined in terms of the Jacobian matrix
\begin{equation}
J(x_0,y_0,t_1) = \begin{pmatrix}
  \partial f_x/\partial x_0 & \partial f_x/\partial y_0\\
  \partial f_y/\partial x_0 & \partial f_y/\partial y_0
\end{pmatrix}\equiv
\begin{pmatrix}
a & b\\
c & d
\end{pmatrix}.
\end{equation}

\subsection{Quasi-separatrix Layers}

The squashing factor at a point $(x_0,y_0)$ of the mapping from time $t_0$ to $t_1$ is
\begin{equation}
Q(x_0,y_0,t_1) = \frac{T}{|D|},
\label{eqn:q}
\end{equation}
where $T = \textrm{Tr}(J^TJ) \equiv a^2+b^2+c^2+d^2$ and $D = \textrm{det}(J) \equiv ad - bc$.
It is a dimensionless measure of the local degree of stretching and squashing of a magnetic flux tube under the field line mapping \citep{titov2002}. Large values of $Q$ signify the locations of strong gradients in the mapping. We find computation of $Q$ to be more robust using the equivalent form
\begin{equation}
Q = R + \frac{1}{R},
\label{eqn:qr}
\end{equation}
where $R=\sqrt{\lambda_+/\lambda_-}$ is the ratio of singular values of $J$ \citep[i.e., $\lambda_\pm$ are the eigenvalues of the positive-definite matrix $J^TJ$;][]{titov2002,richardson2011}.

\begin{figure*}
\sidecaption
\includegraphics[width=12cm]{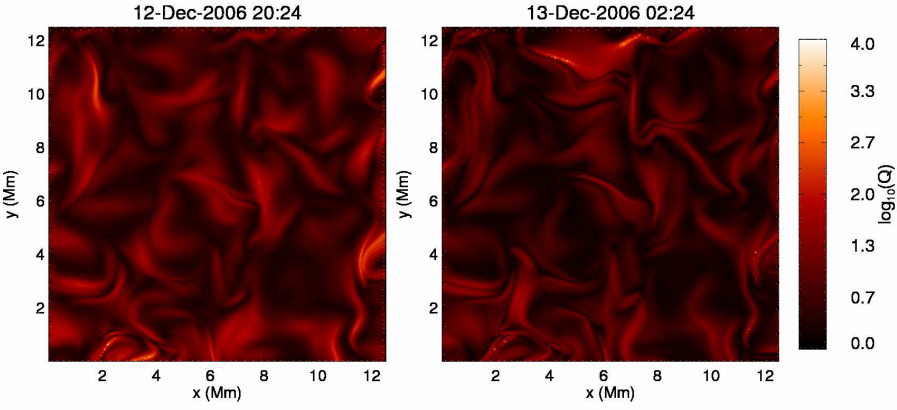}
\caption{Contours of $\log_{10}Q$ after six and twelve hours (at 20:24UT and 02:24UT respectively), with initial time $t_0$ at 14:04UT (available as a movie).}
\label{fig:qobs}
\end{figure*}

\begin{figure}
\centering
\includegraphics[width=\columnwidth]{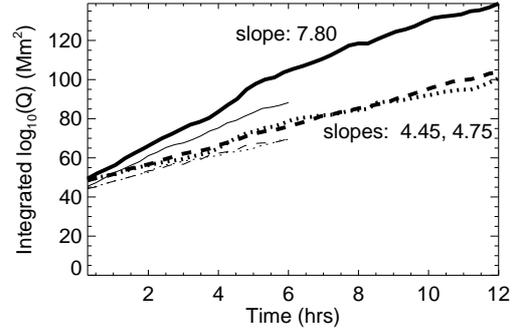}
\caption{Integrated squashing factor, $\int\log_{10}Q\,dxdy$ as a function of time for the observed velocity field (thick lines) and analytical models of Sect. \ref{sec:interp} (thin lines). The solid lines correspond to the observed ${\bf v}$ and combined model 3. Dotted and dashed lines correspond respectively to ${\bf v}_{\rm div}$ (or model 1) and ${\bf v}_{\rm rot}$ (or model 2). The given slopes are for a least-squares linear fit to the observed data.}
\label{fig:qtime}
\end{figure}

We have computed $Q$ as a function of $t_1$ using the inferred field line mapping, and the resulting distribution of $\log_{10}Q$ is shown in Fig. \ref{fig:qobs} at two different times. The logarithmic scaling is introduced because certain trajectories typically become exponentially separated in time: these are precisely the locations of strong gradients in the resulting field line mapping. These thin ``ridges'' of high $Q$, usually known as quasi-separatrix layers (QSLs), are visible at a number of locations in the 12-hour snapshot. They are interspersed with more diffuse regions of $Q$. As a simple measure of the overall structure, we show the time variation of $\int \log_{10}Q\,dxdy$ as the thick, solid curve in Fig. \ref{fig:qtime}. This shows an approximately linear increase through the twelve hours of observations.

All of our figures are shown in the ``initial'' frame $(x_0,y_0)$, so that QSLs correspond to unstable manifolds \citep{richardson2011}. By definition, $Q$ is independent of the direction of mapping along a particular field line, but when calculated for the inverse mapping and plotted at $\big(f_x(x_0,y_0,t_1),f_y(x_0,y_0,t_1) \big)$, the pattern would differ and the apparent QSLs would then correspond to stable manifolds. These stable manifolds tend to be visible observationally because they form the ``network'' along which magnetic flux concentrates over time. This effect is visible in Figure \ref{fig:suspend}. Conversely, we would expect an anti-correlation between the pattern of photospheric magnetic flux at time $t_1$ and the locations of QSLs in Figure \ref{fig:qobs}, which correspond to unstable manifolds.

\subsection{Lagrangian Coherent Structures}

\begin{figure*}
\sidecaption
\includegraphics[width=12cm]{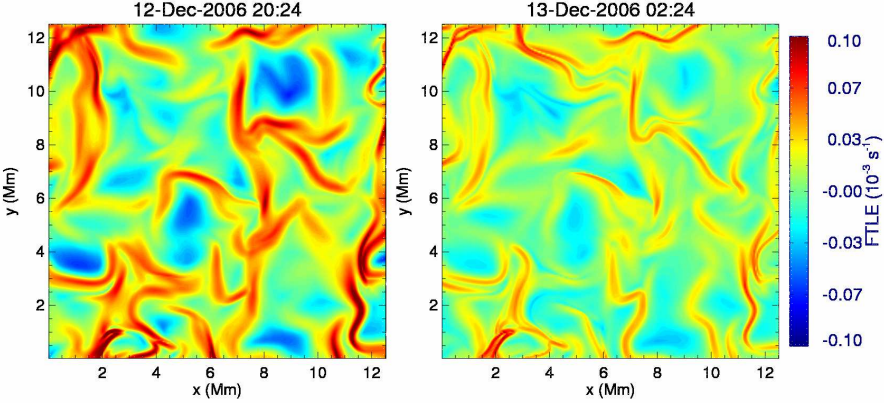}
\caption{Finite-time Lypanunov exponent (FTLE) field $\sigma$ at times 20:24UT and 02:24UT (available as a movie).}
\label{fig:ftle}
\end{figure*}

An alternative measure of the mapping gradient is given by the finite-time Lyapunov exponent or FTLE \citep{haller2001,shadden2005,shadden2011}, defined as
\begin{equation}
\sigma(x_0,y_0,t_1) = \frac{\ln\sqrt{\lambda_+}}{|t_1-t_0|},
\label{eqn:sigma}
\end{equation}
where as before $\lambda_+$ is the largest eigenvalue of $J^TJ$. The quantity $\sigma$ gives the separation rate between two initially nearby trajectories: if two points are initially separated by a small distance ${|\boldsymbol\xi}_0|$ at time $t_0$, then their separation at a later time $t_1$ will be $|{\boldsymbol\xi}_1| \approx \exp[\sigma(t_1-t_0)]|{\boldsymbol\xi}_0|$. We shall use only the forward time FTLE, although one may define also a backward FTLE using the inverse mapping.

In fact, comparing the definition of $\sigma$ to that of $Q$ in Equation \eqref{eqn:q} shows that both quantities measure essentially the norm of the matrix $J^TJ$ (which is the Cauchy-Green tensor of the deformation imposed by the field-line mapping). The quantity $Q$ uses the Frobenius norm of the matrix, whereas $\sigma$ uses the spectral norm. They differ further in that $\sigma$ includes the logarithm in its definition, and has units of inverse time, whereas $Q$ is dimensionless. In the particular case that $D=1$, it follows, in the high-$Q$ layers, that $Q\approx |\lambda_+|$, so that we may make the direct correspondence $\ln Q \approx 2|t_1-t_0|\sigma$.

The distribution of $\sigma$ is shown in Fig. \ref{fig:ftle}, for the same two times $t_1$ as Fig. \ref{fig:qobs}. We see the development of thin ridges of large (positive) $\sigma$, which represent the locally strongest repelling material surfaces in the flow. These are interspersed by regions of negative $\sigma$, indicating converging trajectories. The ridges in Fig. \ref{fig:ftle} coincide precisely with the quasi-separatrix layers in Fig. \ref{fig:qobs}; in the fluid dynamical context they are known as Lagrangian Coherent Structures \citep{haller2000,shadden2005}. They are termed ``lagrangian'' because they are defined by the fluid motion rather than an instantaneous snapshot, and ``coherent'' because they have distinguished stability compared to other nearby material surfaces. Analysis of such structures has become an important tool in fluid dynamics because they often reveal the mechanisms underlying transport in complex fluid flows: for example, the patterns traced out by visual markers such as dye. With our assumption of ideal MHD, the coronal magnetic field inherits these topological structures directly from the photospheric flow.

\section{Interpretation} \label{sec:interp}

The spatial patterns of $Q$ and $\sigma$ in Figs. \ref{fig:qobs} and \ref{fig:ftle} reveal the topological structure of the coronal magnetic field generated by the observed photospheric velocities. We can understand the origin of this structure using a simple analytical model of two-dimensional convection, similar to that of \citet{simon1989}. We will demonstrate the effects of both diverging and rotational flows, before using the model to predict how the inferred field line mapping might differ using higher-resolution velocity observations.

\subsection{Analytical Model}

We present three models, each computed on a grid of size $12\,\textrm{Mm}\times 12\,\textrm{Mm}$.

\begin{figure*}
\centering
\includegraphics[width=17cm]{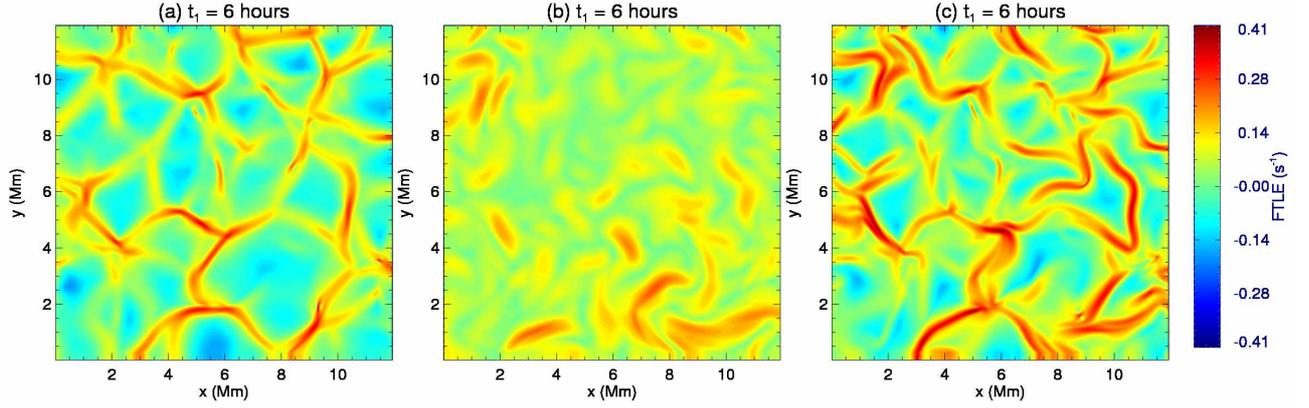}
\caption{Finite-time Lyapunov exponent (FTLE) field $\sigma$ after 6 hours for photospheric flows in the analytical models with (a) diverging flow only, (b) rotational flow only, and (c) both diverging and rotational flow.}
\label{fig:model}
\end{figure*}

\begin{enumerate}
\item{\it Diverging flow}. Here the velocity field ${\bf v}$ is a superposition of convective ``plumes''
\begin{equation}
v_x = \sum_{i=1}^{N} \frac{(x-X_i)}{r_i}v_{ri},\qquad v_y = \sum_{i=1}^{N} \frac{(y-Y_i)}{r_i}v_{ri},
\end{equation}
where $(X_i,Y_i)$ is the centre of the $i^{\rm th}$ plume, $r_i=\sqrt{(x-X_i)^2 + (y-Y_i)^2}$, and the plumes have a purely radial velocity with respect to their centre of
\begin{equation}
v_{ri}=V_i\sqrt{2}\,e^{0.5}\frac{r_i}{R_i}\exp\left(-\frac{r_i^2}{R_i^2} \right).
\end{equation}
Thus $\omega=0$ and $\Delta\neq 0$. The centres $(X_i,Y_i)$ of the $N=81$ plumes are selected at random from a uniform distribution, over a larger domain $-3\le X_i\le 15$,  $-3\le Y_i\le 15$ to avoid boundary effects. To approximate the observations, the peak velocities $V_i$ are selected from a normal distribution with mean $\langle V_i \rangle = 0.1\,\textrm{km}\,{s}^{-1}$ and standard deviation ${\rm sd}(V_i)=0.03\,\textrm{km}\,{s}^{-1}$, while the widths $R_i$ are selected from a normal distribution $\langle R_i \rangle =1\,\textrm{Mm}$ and ${\rm sd}(R_i)=0.3\,\textrm{Mm}$. For computational convenience, the same pattern of plumes is used for a ``coherence time'' of $15$ minutes, before a new pattern is chosen. (Qualitatively similar results are obtained if plumes are given a Gaussian profile in time, with randomly distributed peak times.) The choice of $15$ minutes is consistent with the observations of \citet{welsch2011} for flows on this spatial scale.

\item{\it Rotational flow}. Here the velocity field ${\bf v}$ is a superposition of vortices 
\begin{equation}
v_x = \sum_{i=1}^{\tilde{N}} -\frac{(y-\tilde{Y_i})}{\tilde{r}_i}\tilde{v}_{\phi i},\qquad v_y = \sum_{i=1}^{\tilde{N}} \frac{(x-\tilde{X}_i)}{\tilde{r}_i}\tilde{v}_{\phi i},
\end{equation}
where $(\tilde{X}_i,\tilde{Y}_i)$ is the centre of the $i^{\rm th}$ vortex, $\tilde{r}_i=\sqrt{(x-\tilde{X}_i)^2 + (y-\tilde{Y}_i)^2}$, and the vortex has an azimuthal velocity
\begin{equation}
\tilde{v}_{\phi i}=\frac{\tilde{\Omega}_ir_i}{2}\exp\left(-\frac{\tilde{r}_i^2}{\tilde{R}_i^2} \right).
\end{equation}
Now $\omega\neq 0$ and $\Delta=0$. We choose $\tilde{N}=81$ vortices, with centres $(\tilde{X}_i,\tilde{Y}_i)$ on an extended grid as in model 1. Since $\omega$ and $\Delta$ are observed to have comparable scales (Sect. \ref{sec:data}), the widths $\tilde{R}_i$ are again selected from a normal distribution with $\langle \tilde{R}_i\rangle =1\,\textrm{Mm}$ and ${\rm sd}(\tilde{R}_i)=0.3\,\textrm{Mm}$. The strengths $\tilde{\Omega}_i$ are selected from a normal distribution with zero mean and ${\rm sd}(\tilde{\Omega}_i)=0.00048\,\textrm{s}^{-1}$. The latter value makes the RMS $\omega$ in the box approximately equal to the RMS $\Delta$ in model 1 (approximately $2\times 10^{-4}\,\textrm{s}^{-1}$). Again, a constant pattern of vortices is applied for each coherence time.

\item {\it Diverging and rotational flow}. The velocity field is the superposition of those from models 1 and 2.
\end{enumerate}
In each model, an initial grid of  $864\times 864$ tracer points has been integrated in the same way as for the observational data, following trajectories of the model flows. The resulting $\sigma$ fields for each model after 6 hours (corresponding to 24 coherence times) are shown in Fig. \ref{fig:model}a--c. The time evolution of $\int\log_{10}Q\,dxdy$ is shown by the thin lines in Fig. \ref{fig:qtime} for models 1 (dotted), 2 (dashed) and 3 (solid). Note that the combined model 3 leads to a linear increase at a rate comparable to the observed flow (thick solid line). This rate is approximately double that of models 1 or 2.

\subsection{Origin of the Observed Pattern}

From the $\sigma$ fields for the three analytical models in Fig. \ref{fig:model}, we can see that the qualitative structure of the \emph{observed} $\sigma$ field (Fig. \ref{fig:ftle}) reflects predominantly the influence of divergence $\Delta$ in the velocity field. This is responsible for generating the visible ridge network of LCSs (or QSLs), interspersed with regions where $\sigma < 0$. Note that this network differs from that of the magnetic flux distribution (Fig. \ref{fig:suspend}): the latter collects along the stable manifolds rather than the unstable manifolds.

Model 2, with purely rotational flow (Fig. \ref{fig:model}b), generates a very different $\sigma$ field: there are still localised ridges, but these are more diffuse and space-filling than in the diverging flow. There are no regions of $\sigma < 0$. When both flows are superimposed (Fig. \ref{fig:model}c), the $\sigma$ field visually resembles that of the diverging flow, with the contribution of vorticity being to locally perturb the ridge network and to infill some of the $\sigma<0$ regions with new ridges (although, in fact, $\int\log_{10}Q\,dxdy$ increases at twice the rate of model 1). This model, where $\Delta$ and $\omega$ are comparable in magnitude, corresponds most closely with the observed velocity field.

Although rotational component of the combined flow has little visual effect on the LCS/QSL pattern, this does not mean that it is incapable of leading to gradients and subsequent reconnection in the coronal magnetic field. Rather, the superposition of a diverging flow has perturbed the picture of the $\sigma$ and $Q$ fields so as to mask the contribution from $\omega$. It may therefore prove useful for future analysis to extract the two components from the observed velocity field. This may be done through a Helmholtz decomposition
\begin{equation}
{\bf v} = {\bf v}_{\rm div} + {\bf v}_{\rm rot},
\end{equation}
where ${\bf v}_{\rm div}=\nabla\phi$ and ${\bf v}_{\rm rot}=\nabla\times(\psi{\bf e}_z)$, for scalar functions $\phi(x,y,t)$, $\psi(x,y,t)$. This decomposition is not unique but we can fix a particular solution by specifying the boundary conditions that ${\bf n}\cdot{\bf v}_{\rm rot}=0$ on the boundary of our square region, where ${\bf n}$ is the unit normal to the boundary. Then ${\bf v}\cdot{\bf n}={\bf v}_{\rm div}\cdot{\bf n}$ on the boundary. The function $\phi$ is determined by solving the Poisson equation $\nabla^2\phi = \Delta$ with the Neumann boundary conditions that ${\bf n}\cdot\nabla\phi = {\bf n}\cdot{\bf v}$ on the boundary. The function $\psi$ is determined by solving the Poisson equation
\begin{equation}
\nabla^2\psi = -\omega,
\label{eqn:poisson}
\end{equation}
in this case with the Dirichlet condition of constant $\psi$ on the boundary.

\begin{figure*}
\sidecaption
\includegraphics[width=12cm]{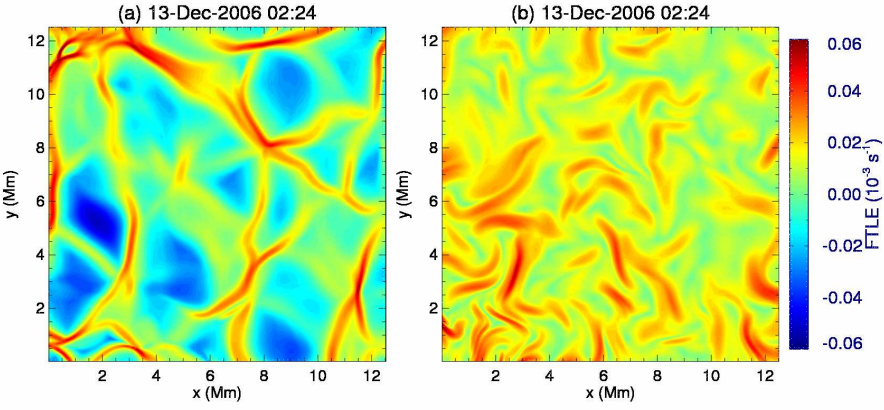}
\caption{Finite-time Lyapunov exponent (FTLE) field $\sigma$ at 02:24UT for the (a) diverging component ${\bf v}_{\rm div}$ and (b) the rotational component ${\bf v}_{\rm rot}$ of the Helmholtz-decomposed observed velocity field.}
\label{fig:decomp}
\end{figure*}

Given $\omega$ from the observed velocity, we calculated $\psi(x,y,t)$ at each time slice by solving Equation \eqref{eqn:poisson} using a standard fast Poisson solver \citep{vanloan1992}. The $\sigma$ fields were then calculated for ${\bf v}_{\rm div}$ and ${\bf v}_{\rm rot}$ separately, and are shown at 02:24UT in Fig. \ref{fig:decomp}. This confirms our qualitative picture from the analytical model of the rather different $\sigma$ fields for the two types of flow. The thick dotted and dashed curves in Fig. \ref{fig:qtime} show that $\int\log_{10}Q\,dxdy$ for both ${\bf v}_{\rm div}$ and ${\bf v}_{\rm rot}$ increases at about half the rate of that for the combined velocity, again in accordance with the model.

\subsection{Predicted Structure from Faster Flows}

Since our method relies on the photospheric velocity field ${\bf v}(x,y,t)$ as input, the inferred magnetic structure is necessarily affected by limitations in the velocity observations used. In addition to the limited spatial resolution of SOT (which has a diffraction limit of approximately $0.3$''), the velocities we report have been averaged in both space and time in order to reduce noise in the local correlation tracking. Previous studies of granular convection on the Sun report cells of size $0.5-2\,\textrm{Mm}$ with typical lifetimes of $\sim10$ minutes, and velocities from $0.5-1.5\,\textrm{km}\,\textrm{s}^{-1}$ \citep{rieutord2010}. While the ``cells'' we find are comparable to the larger end of this size range, our velocities (Fig. \ref{fig:location}b) are a factor of $5-10$ smaller due to the effect of the averaging.

\begin{figure*}
\centering
\includegraphics[width=17cm]{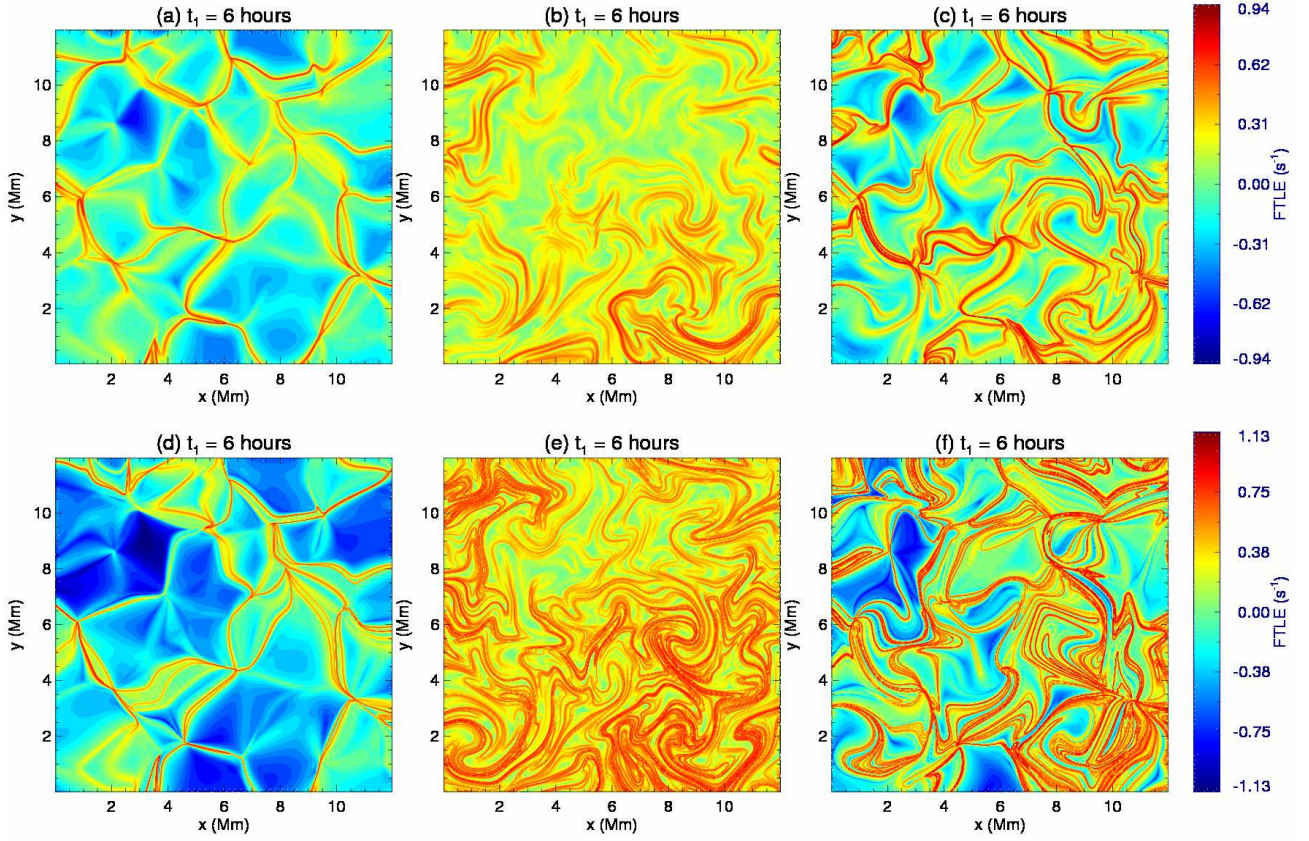}
\caption{Effect of an increased flow speed in the analytical model. In panels (a)--(c) the mean plume velocity is $0.3\,\textrm{km}\,\textrm{s}^{-1}$, while in panels (d)--(f) it is $0.5\,\textrm{km}\,\textrm{s}^{-1}$. As in Fig. \ref{fig:model}, the $\sigma$ field is shown after 6 hours for models 1 (left column), 2 (middle), and 3 (right).}
\label{fig:faster}
\end{figure*}

To predict the magnetic structure that would arise from faster flows, we re-run the analytical model with plume velocity increased first to $\langle V_i\rangle=0.3\,\textrm{km}\,\textrm{s}^{-1}$ and then to $\langle V_i\rangle=0.5\,\textrm{km}\,\textrm{s}^{-1}$. To keep the RMS vorticity comparable to the RMS divergence, we also increase ${\rm sd}(\tilde{\Omega}_i)$ to $0.00144\,\textrm{s}^{-1}$ and $0.0024\,\textrm{s}^{-1}$ respectively. The resulting $\sigma$ fields in models 1, 2 and 3 after 6 hours are shown in Fig. \ref{fig:faster}. The results are striking: in model 1 (diverging flow), the effect of faster flows is to sharpen the LCSs, increasing the maximum value of $\sigma$ but maintaining the overall pattern. By contrast, in model 2 (rotational flow), faster flows lead not only to sharper LCSs but also to greater filling of space with these structures. Interestingly, model 3 (combined flows) sees not only the sharpening effect of model 1 but also an increased infilling of the $\sigma<0$ regions with new LCSs resulting from the vortices.

The infilling arises because, with strong enough vorticity, LCSs ``wind up'' around centres in the flow pattern. This is particularly evident in regions where the vorticity is strong but the divergence is weak; for example, compare the region around $x=1$, $y=1.5$ in Figs. \ref{fig:faster}d-f. This ``wind up'' phenomenon was found by \citet{demoulin1996a} who computed QSLs in analytical flux tubes of increasing twist. Similarly, \citet{birn1989} found that field line connectivity in a 3D plasmoid varies on smaller and smaller scales as the axial field is reduced relative to the azimuthal field, with the boundaries between regions of different connectivity wrapping round and round the flux tube axis.

The initial rate of increase in $\int\log_{10}Q\,dxdy$ (corresponding to the slopes in Fig. \ref{fig:qtime}) increases from $\sim 8$ when $\langle V_i\rangle=0.1\,\textrm{km}\,\textrm{s}^{-1}$  to $\sim 44$ when $\langle V_i\rangle=0.3\,\textrm{km}\,\textrm{s}^{-1}$ and $\sim 91$ when $\langle V_i\rangle=0.5\,\textrm{km}\,\textrm{s}^{-1}$, although the rate of increase slows considerably in the latter case during the 6 hours. This may correspond to a ``saturation'' of the infilling evident in models 2 and 3 once the $\sigma$ field has become homogenised, in accordance with previous studies of LCSs in turbulent convection \citep{lapeyre2002}. At this stage, magnetic field lines in the whole region will have become mixed/braided in a manner likely to promote reconnection and subsequently heating of the coronal plasma. A saturation is also seen in the maximum value of $\sigma$, and this occurs after a shorter time for higher flow speed. This may be due to the LCS widths falling below the tracing grid scale, but it is unclear what other effects might cause such saturation; this bears further investigation.

\begin{figure*}
\centering
\includegraphics[width=17cm]{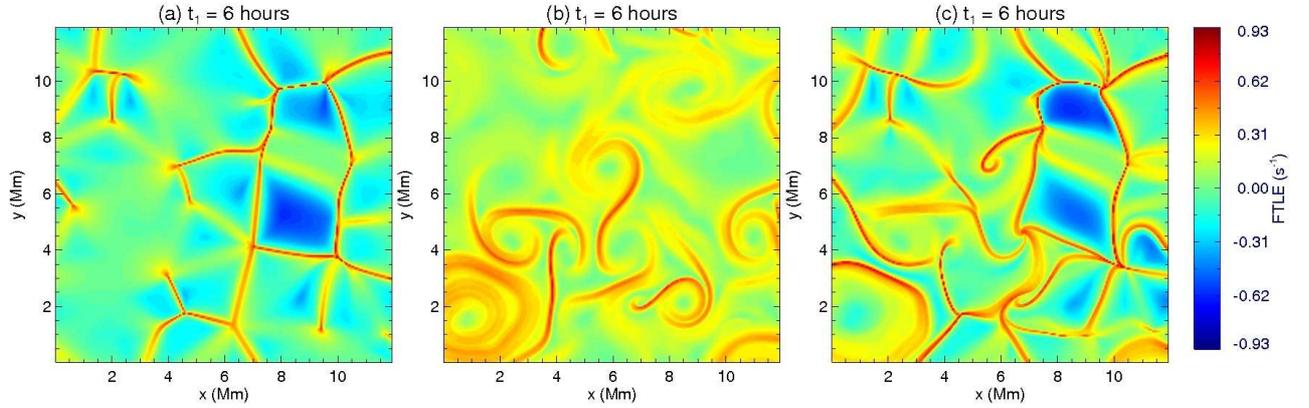}
\caption{Finite-time Lyapunov exponent (FTLE) field $\sigma$ after 6 hours of a stationary velocity field in the analytical models with (a) diverging flow only, (b) rotational flow only, and (c) both diverging and rotational flow.}
\label{fig:long}
\end{figure*}

Note that certain consequences of a faster plume velocity, namely (i) sharper LCS with higher peaks of $\sigma$, and (ii) a faster rate of increase in $\int\log_{10}Q\,dxdy$, would also result if one left the plume velocities unchanged but increased their coherence time. This is demonstrated in Fig. \ref{fig:long}, which shows model results after six hours with a single velocity pattern, rather than changing the pattern every 15 minutes. The peak $\sigma$ (c.f. Fig. \ref{fig:model}) and slope of the $\int\log_{10}Q\,dxdy$ curve (not shown) become comparable to the run with $\langle V_i\rangle=0.3\,\textrm{km}\,\textrm{s}^{-1}$. Yet there are fewer LCS, filling less of the area. This illustrates how the pattern of $\sigma$ depends on the time history of the flow, not just on its pattern at any given instant.

For simplicity, the models here are limited to convective cells with a single spatial scale, flow speed and lifetime. On the real Sun, convection seems to operate simultaneously on a range of scales. Experiments with superimposing a ``supergranular'' flow in the model (cells 10 times larger, with slower speeds and longer lifetimes) indicate that the LCS pattern and increase of $\int\log_{10}Q\,dxdy$ are determined primarily by the original, faster flow component. The supergranular flow can generate localised magnetic gradients only over a longer timescale of many hours.

\section{Conclusion} \label{sec:conclusions}

We have proposed a practical method for inferring the topology of the 3D coronal magnetic field not by extrapolation (as is typically used) but rather by integrating trajectories of an observed sequence of horizontal flows in the photosphere. Assuming an ideal evolution in the corona, this enables us to infer the resulting magnetic field line mapping between photospheric footpoints, and therefore the squashing factor $Q$ of this mapping. The method has been demonstrated using a particular sequence of observed flows derived from Hinode SOT magnetograms, although it applies equally to any velocity data and does not require magnetic information. Since the inferred field line mapping results directly from the observed photospheric flow, quasi-separatrix layers in the coronal magnetic field correspond to so-called Lagrangian Coherent Structures in the flow. These are ridges of high finite-time Lyapunov exponent $\sigma$ where neighbouring trajectories diverge most strongly.

With a simple analytical model, we have demonstrated that the pattern of the $\sigma$ (or equivalently $Q$) field differs significantly between a flow composed of irrotational convective plumes and one composed of purely incompressible vortices. The diverging flow pattern in the first case leads to a network of long thin LCSs, whereas the vortical flow leads to a space-filling pattern of shorter LCSs that look \emph{a priori} more favourable for widespread reconnection. The observed velocity field is found to have comparable $|\nabla\cdot{\bf v}|$ and $|\nabla\times{\bf v}|$, and the observed $\sigma$ field is in qualitative agreement with a combined model incorporating both effects. In the combined case, the appearance of the $\sigma$ and $Q$ fields follows that of the diverging flow model: the diverging part of the velocity acts to quickly stretch and deform the picture. However, the rate of increase of integrated $\log_{10}Q$ in the combined model is double that of the original model, and we hypothesise that the vortical structure remains ``hidden'' in the magnetic field topology. We have demonstrated how the vortical part may be extracted from an observed velocity field, but further study is required to determine whether the vortical part is a more appropriate predictor of subsequent reconnection.

Due to the limitations of the observational technique used for the demonstration in this paper, the typical flow speeds measured were only $0.1\,\textrm{km}\,\textrm{s}^{-1}$, a factor of $5-10$ slower than real granular flows. From investigation of the analytical model, we predict that faster flow speeds (for the same size and lifetime of granules) will result in significantly faster development of strong gradients in the magnetic field. Our initial results are therefore very much a lower bound for the complexity that we expect to develop in the coronal magnetic field over this time. The model also indicates that, if the real vorticity is also larger, then the combined $\sigma$ field will show greater infilling of LCSs. In this case, the mixing of trajectories is sufficient that the model begins to show a process of ``homogenisation'' of the $\sigma$ field as found in simulations of turbulence \citep{lapeyre2002}. A proper investigation of the rate of this mixing will require higher resolution velocity data, but the simple model indicates that it is likely to be effective on timescales much shorter than the observed 12 hours. Realistic numerical simulations of photospheric convection \citep[e.g.,][]{stein1998,gudiksen2005} could also give tighter constraints on the expected magnetic structure.

Finally, the method proposed here---which assumes a perfectly ideal evolution in the corona---will break down if and when sufficiently high magnetic gradients have formed for magnetic reconnection to set in. Determining this threshold for reconnection will likely require detailed study of numerical MHD simulations. To this end, we have proposed a method for reconstructing a 3D magnetic field with the inferred field line mapping. This field is neither unique nor (likely) in equilibrium, but it has the significant advantage over existing extrapolation techniques of having the correct field line topology.

\begin{acknowledgements}
ARY and GH were supported by the UK STFC (grant ST/G002436/1) to the University of Dundee. \emph{Hinode} is a Japanese mission developed and launched by ISAS/JAXA, with NAOJ as domestic partner and NASA and STFC (UK) as international partners. It is operated by these agencies in co-operation with ESA and NSC (Norway).
\end{acknowledgements}

\bibliographystyle{aa}
\bibliography{yeatesAA201118278}

\appendix

\section{Magnetic Field Construction} \label{sec:mag}

The method outlined in this paper infers the magnetic field line mapping from a photospheric flow. However, the field line mapping defines the 3D magnetic field ${\bf B}$ only up to an ideal deformation. In particular, a hypothetical plasma flow in the volume, which vanishes on the photospheric boundaries, can deform ${\bf B}$ while leaving the field line mapping invariant. Conversely, two fields ${\bf B}$ resulting from applying the same photospheric footpoint motions to the same initial magnetic field can differ at most by an ideal deformation. 

It would be useful for future investigations to generate a particular magnetic field of the required topology. Here we present a general method for constructing such a field. Given a velocity field ${\bf v}(x,y,t)$, the strategy is to set
\begin{equation}
{\bf B}(x,y,z) = \lambda(x,y,z)\Big(v_x(x,y,z){\bf e}_x + v_y(x,y,z){\bf e}_y + {\bf e}_z \Big)
\end{equation}
so that the field lines of ${\bf B}$ are simply the trajectories of ${\bf v}$ with $z$ corresponding to time. The scalar function $\lambda$ is then adjusted to make ${\bf B}$ divergence-free and match a given normal component  $B_z(x,y,0)$ on the lower boundary. From $\nabla\cdot{\bf B}=0$ we find
\begin{equation}
\frac{d\lambda}{dz} = -\lambda\Delta,
\end{equation}
where the $z$-derivative is taken along a magnetic field line and $\Delta\equiv\nabla\cdot{\bf v}$ as before. Integrating from $t=0$ to $t=z$ along the field line ${\bf f}(x_0,y_0,t)$ gives
\begin{equation}
\int_{B_z(x_0,y_0,0)}^\lambda \frac{d\lambda}{\lambda} = -\int_0^z\Delta\Big({\bf f}(x_0,y_0,t)\Big)\,dt,
\end{equation}
so at the point $(x,y,z)={\bf f}(x_0,y_0,z)$ we can integrate to find
\begin{equation}
\lambda(x,y,z) = B_z(x_0,y_0,0)\exp\left(-\int_0^z\Delta\Big({\bf f}(x_0,y_0,t)\Big)\,dt \right).
\end{equation}
Notice that the magnetic field is entirely determined by knowing both its magnetic field lines and the distribution of $B_z$ on the lower boundary. The magnetic field in Fig. \ref{fig:suspend} was constructed in this way from the observed velocity field.

\end{document}